\documentclass[preprint,aps]{revtex4}

\usepackage{latexsym}
\usepackage{graphicx}
\usepackage{amsfonts}
\usepackage{amssymb}
\usepackage{amsmath}
\usepackage{amsbsy}
\usepackage[english,]{babel}
\usepackage{natbib}

\newcommand{\vect}[1]{{\bf #1}}
\newcommand{\bnabla}{\boldsymbol{\nabla}}

\newcommand{\md}[1]{\mathcal #1}
\newcommand{\mb}[1]{\mathbb #1}

\newcommand{\sech}{{\rm sech}}

\newcommand{\vunit}{\ensuremath{\,\mbox{km} \, \mbox{s}^{-1}}}
\newcommand{\echo}{ECHO }

\newcommand{\dpt}{\partial_t}

\newcommand{\ba}{\begin{eqnarray}}
\newcommand{\ea}{\end{eqnarray}}

\newcommand{\bas}{\begin{eqnarray*}}
\newcommand{\eas}{\end{eqnarray*}}

\newcommand{\be}{\begin{equation}}
\newcommand{\ee}{\end{equation}}

\newcommand{\bes}{\begin{equation*}}
\newcommand{\ees}{\end{equation*}}

\newcommand{\bts}{\begin{figure*}[t!]}
\newcommand{\bhs}{\begin{figure*}[h!]}
\newcommand{\bbs}{\begin{figure*}[b!]}
\newcommand{\efs}{\end{figure*}}
\newcommand{\bfd}{\begin{figure}[tbh]}
\newcommand{\bft}{\begin{figure}[t!]}
\newcommand{\bfh}{\begin{figure}[h!]}
\newcommand{\bfb}{\begin{figure}[b!]}
\newcommand{\ef}{\end{figure}}

\newcommand{\bd}{\begin{displaymath}}
\newcommand{\ed}{\end{displaymath}}

\begin{document}

\title{Three-dimensional evolution of magnetic and velocity shear driven instabilities in a compressible magnetized jet}

\author{Lapo Bettarini}
\email{lapo.bettarini@wis.kuleuven.be}
\affiliation{
Katholieke Universiteit Leuven, Centrum voor Plasma Astrofysica, Celestijnenlaan 200B, B-3001, Leuven, Belgium\\
Dipartimento di Astronomia e Scienza dello Spazio, Universit\`a degli Studi di Firenze, Largo E. Fermi, 2, I-50125, Firenze, Italy}
\author{Simone Landi}
\affiliation{
Dipartimento di Astronomia e Scienza dello Spazio, Universit\`a degli Studi di Firenze, Largo E. Fermi, 2, I-50125, Firenze, Italy}
\author{Marco Velli}
\affiliation{
Jet Propulsion Laboratory, 4800 Oak Grove Drive, Pasadena, CA-91109, USA\\
Dipartimento di Astronomia e Scienza dello Spazio, Universit\`a degli Studi di Firenze, Largo E. Fermi, 2, I-50125, Firenze, Italy}
\author{Pasquale Londrillo}
\affiliation{
INAF  Osservatorio Astronomico di Bologna, via C. Ranzani 1, I-40127 Bologna, Italy}

\begin{abstract}
The problem of three-dimensional combined magnetic and velocity shear driven instabilities of a compressible magnetized jet modeled with a plane neutral/current double vortex sheet in the framework of the resistive magnetohydrodynamics is addressed. The resulting dynamics given by the stream$+$current sheet interaction is analyzed and the effects of a variable geometry of the basic fields are considered. Depending on the basic asymptotic magnetic field configuration, a selection rule of the linear instability modes can be obtained. Hence, the system follows a two-stage path developing either through a fully three-dimensional dynamics with a rapid evolution of kink modes leading to a final turbulent state, or rather through a driving two-dimensional instability pattern that develops on parallel planes on which a reconnection$+$coalescence process takes place. 
\end{abstract}

\maketitle

\section{Introduction} \label{sec1}
Velocity shears and strong magnetic field gradients are known to play an important role in the dynamics of both laboratory and astrophysical plasmas, such as solar flares~\citep{ofman91,tsuneta96}, earth's magnetotail~\citep{uberoi84,pu90}, tokamaks~\citep{paris83}, and, in the inner heliosphere, the interaction of the heliospheric current-sheet with the structure determined by the slow component of the solar wind on the solar equatorial plane, embedded in the fast component~\citep{einaudi99,einaudi01,rappazzo05,bettarini06}. In fact, in a reference frame co-moving with the slow wind, we have a bimodal flow profile where the velocity is zero at the HCS, and across it the interplanetary magnetic field (IMF) changes sign from the Southern to the Northern solar hemispheres, regions of fast wind, resulting in a wake flow profile whose three-dimensional evolution is not fully understood.

The incompressible two- and three-dimensional instability dynamics of a plane current vortex sheet has been largely investigated in previous works~\citep{dahlburg97,dahlburg01} and only recently compressibility effects have started to be accounted at least for the linear  regime~\citep{dahlburg00}. In~\citet{dahlburg97}, the simplest two-dimensional incompressible system is analyzed in the linear and nonlinear regime: The basic magnetic field and velocity field are both modeled by a hyperbolic tangent profile with a variable ratio of the velocity to the magnetic shear width, $\delta$, for different values of the Alfv\`enic Mach number, $\md{M}_a$. The instability behavior of such systems depends strongly on these two parameters. In particular, for small values of $\md{M}_a$ and large values of $\delta$, a direct energy transfer from the basic velocity field to the perturbed magnetic field is observed and, in general, the instability evolution can not be considered simply as a mixture of the Kelvin-Helmholtz (KH, hereafter) and resistive instabilities. 
\bfb
\begin{center}
\includegraphics[width=0.4\textwidth]{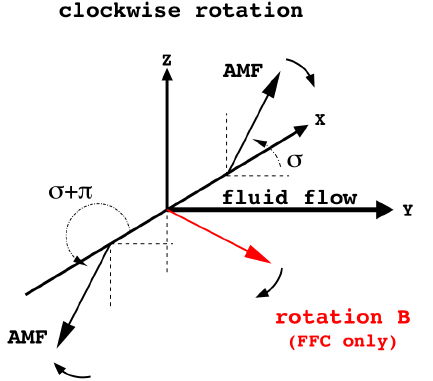}
\caption{(Color online) Schematic reconstruction of the current/neutral double vortex-sheet model used in the present work. The main features of the generic configuration are shown: the direction of the AMF, the type of rotation for the FF magnetic field (clockwise or counterclockwise with respect to the rotation axis, $x$), and the presence (absence) of the a magnetic field component within the fluid flow for the FF (PB).}
\label{fig1}
\end{center}
\ef
The presence of a third direction is expected to modify the linear and the nonlinear regime. First, it allows the ignition of all the instability modes of the system and, in particular, all the modes that are orthogonal to the basic flow; secondly, it allows the system to go from an initially laminar state to a turbulent one by means of a three-step process (a primary instability, e. g. the tearing mode for the neutral sheet; a two-dimensional quasi-steady state; a further "secondary instability", caused by three-dimensional disturbances). Such processes have been observed in several hydrodynamic and magnetohydrodynamic (MHD) configurations such as vortex~\citep{metcalfe87} and neutral sheets~\citep{dahlburg92,schumacher00}. \citet{dahlburg01} consider an incompressible problem with a subsonic shear flow whose profile is given by a hyperbolic tangent function with a width equal to that one of the neutral sheet. The perturbations they use consist of two-dimensional disturbances of large amplitude, so that the system should be near the saturation condition in the nonlinear regime, with the addition of small amplitude three-dimensional modes. The nonlinear evolution of those systems produces a complex structure wherein the three-dimensional effects dominate with respect the two-dimensional evolution. The onset of the secondary instability has consequences also in the wake/jet dynamics. During the nonlinear stage of the resistive primary instability, the formation of magnetically confined plasma structures (plasmoids), is accompanied by the acceleration/deceleration of the central portion of the wake/jet~\citep{einaudi99,einaudi01}. Once the secondary-instability sets in, the free energy of the velocity shear triggers the transition towards a turbulent state and the acceleration/deceleration stops~\citep{einaudi99}.

Compressive effects have important consequences on the linear regime of the plane current-vortex sheet. Weakly evanescent modes are observed to grow and act also at large distances from the shear layers. Also, as shown in recent works on the three-dimensional evolution of the compressive tearing instability~\citep{onofri04,landi08}, the onset of the secondary instability is observed to be critically dependent on the nature of the equilibrium, e.g. whether it is force-free or pressure-balanced, with or without a guide field.

If we consider a more complex two-dimensional system given by a jet/wake interacting with a current/neutral double vortex sheet in a compressible and magnetically-dominated situation~\citep{bettarini06}, the dynamics is observed to be dependent by the relative configuration of the basic magnetic to the velocity field, that is, as sketched in Fig.~\ref{fig1}, by the angle $\sigma$ between the initial jet flow (or equivalently, the wake flow) direction and the asymptotic magnetic field (AMF, hereafter), that is the magnetic field ``far enough'' from the stream center. In particular, when $\sigma = \pi/2$ in low beta configurations of supersonic, but subalfv\'enic shear flows (so, the AMF is orthogonal to the basic jet/wake flow) a typical KH instability fully develops and dominates the dynamics in the non linear regime. For any other angle in between $\sigma = 0$ (the AMF being aligned to the jet/wake) and $\sigma < \pi/2$, a linear varicose-resistive instability~\citep{Dahlburg98} is dominating leading to magnetic field islands which afterwards coalesce. In the present work, we extend the previous analysis to the three-dimensional case both in the linear and nonlinear regime.

In the following sections, we introduce the numerical setup, the initial conditions, the perturbations, and the parameters for our analysis (Sect.~\ref{sec2}). In Sect~\ref{sec3}, we present the results of the simulations for the several different configurations under consideration. We will show how the details of the linear instability dynamics depend on the system's large-scale structure, in particular on the AMF's direction, regardless of the type of equilibrium considered and how this drives into different dynamical paths in the nonlinear regime. In Sect.~\ref{sec4}, we summarize the results drawing our conclusions and pointing to still open questions. 

\section{Equations and numerical settings} \label{sec2}

We solve the set of one-fluid compressive-resistive MHD equations in Cartesian geometry. We define a stream-wise, or Fourier, direction ($y$) along which we impose periodic boundary conditions, a cross-stream direction ($x$) along which the mean flow varies and we impose reflecting boundary conditions, and a span-wise direction ($z$), corresponding to an invariance direction for the quantities describing our initial system and along which we impose periodic boundary conditions. 

In a compressible MHD system where high Mach number flows are supposed to form and afterwards to undergo to resistively-triggered instabilities, the numerical strategy is a critical choice in order to obtain the best results in tracking the physical details of the expected dynamics. On the one hand, the understanding of dissipative processes like magnetic reconnection requires high-order spectral-like numerical techniques such that the physical magnetic diffusion can be fully  appreciated against the numerical
dissipation. Furthermore, high-order methods allow to track the energy spectrum cascade towards high wave-vectors with a reasonable number of grid points. On the other hand, supersonic flows can arise either in the initial state of the system or as a result of the reconnection process in a low plasma beta regime, and they determine plasma and
magnetic field discontinuities that can be properly treated by means of shock-capturing techniques. 

Here, we use the Eulerian Conservative High-Order code (hereafter, \echo) to solve the full set of compressible and resistive MHD equations in a conservative form within the Upwind Constrained Transport (UCT) framework~\citep{londrillo00,londrillo04} which properly implements the magnetic field divergence-free condition. In general, \echo solves the one-fluid equations for a magnetized plasma in different approximations, ranging from the special and general relativistic MHD frameworks~\citep{ldz03,ldz07} to situations wherein the presence of the magnetic field diffusivity is taken into account~\citep{landi08}. \echo is able to handle different high-order numerical techniques for the flux reconstruction and the computation of the derivatives~\citep{ldz07,landi08}. In particular, it allows the use of compact (or implicit) algorithms~\citep{lele92} which have spectral-like resolution properties better than the corresponding explicit numerical techniques~\citep{landi08}. 

The set of equations to be solved is the following:
\ba
\label{continuity}
\dpt \rho & = & -\bnabla \cdot \left(\rho \, \vect{v} \right) \\
\label{momentum}
\dpt \left(\rho \, \vect{v} \right) & = & -\bnabla \cdot \left(\rho \,  \vect{v} \, \vect{v} + \md{P} \mb{I} - \vect{B} \, \vect{B} \right) \\
\label{energy}
\dpt e & = & -\bnabla \cdot \left[\vect{v} \, \left( e + \md{P} \right) - \vect{B} \, \left( \vect{v} \cdot \vect{B} \right) + \right.\nonumber \\
& & \left. - \eta \, \vect{B} \times \bnabla \times \vect{B} \right] \\
\label{Beq}
\dpt \vect{B} & = & -\bnabla \times \vect{E} \\
\label{Eeq}
\vect{E} & = & -\vect{v} \times \vect{B} + \eta \, \bnabla \times \vect{B}
\ea
being $\rho$ the mass density, $\vect{v}$ the fluid velocity, $e = p/(\gamma -1) + \rho \, |\vect{v}|^2 / 2 + |\vect{B}|^2 / 2$ the total (= hydro \textit{plus} magnetic) energy density according to a perfect gas equation of state, where $p$ is the plasma pressure and $\gamma$ the adiabatic index; $\md{P} = p + |\vect{B}|^2 / 2$ the total pressure. In Eqs.~\eqref{energy} and \eqref{Eeq}, an explicit dissipative term due to the plasma resistivity, $\eta$, is present though maintaining  the same formal structure of the conservative framework.

In order to obtain the dimensionless equations, we use the characteristic quantities $\tilde{L}$, $\tilde{\rho}$, $\tilde{v}_a$, which correspond respectively to the velocity shear width, to the  initial uniform density and to the the Alfv\'en velocity at the cross-stream boundaries. Time, velocity and magnetic field strength are expressed in units of the related quantities $\tilde{t}$, $\tilde{v}$, $\tilde{B}$ 
\bes
\tilde{t} = \frac{\tilde{L}}{\tilde{v}_a} \, , \quad \tilde{v} = \md{M}_a \, \tilde{v}_a \, \quad \tilde{B} = \tilde{v}_a \, \sqrt{4\pi \,\tilde{\rho}}.  
\ees
being $\md{M}_a$ the Alfv\'enic Mach number, as already mentioned in the introduction. Moreover, plasma pressure and magnetic diffusivity are measured in terms of  $\tilde{p}=\beta \,\tilde{B}^2/2$ and of the inverse of the Lundquist number $\md{S}= \tilde{v}_a\tilde{L}/\eta$.

\subsection{Simulation set-up} \label{sec2:1}

In the present work, two different configurations are considered: a force-free (FF, hereafter) and a pressure balance (PB, hereafter) initial equilibrium. As shown in Fig.~\ref{fig1}, the FF current sheet is formed by a rotation of the cross-sheet component of the magnetic field across the flow, whereas in the PB case no rotation is considered and the polarity reversal is obtained by means of a neutral sheet. So, FF and PB cases differ essentially due to presence of the rotation component of the magnetic field along the flow (defined as the stream-wise direction) which is absent for $\sigma = 0$ (being directed along the orthogonal or span-wise direction) and becomes more and more relevant as the angle increases. Furthermore, for $\sigma = 0$ and $\pi/2$, the FF is intrinsically symmetric with respect to the stream-wise direction, while this system does not have a defined symmetry at all for $0 < \sigma < \pi/2$. For the PB configuration, the system is always symmetric and in correspondence of the jet/wake we always have a neutral line. This initial parity property of the two configurations determines the evolution of the current and neutral sheet, since in the FF we have a preferential side for magnetic islands formation and a differential deceleration of the jet: the maximum distortion effect is observed for $\sigma = 3\pi/8$~\citep{bettarini06}. 

The FF basic fields are the following
\ba
\label{FF_v0y}
v_{0y}(x) & = & \md{M}_a \, \sech^2 \left( x \right) \\
\label{FF_b0y}
B_{0y}(x) & = & \left[ \sin\sigma \, \, \sech \left( \delta x \right) + \cos\sigma \, \, \tanh \left( \delta x \right) \right] \\
 \label{FF_b0z}
B_{0z}(x) & = & \left[-\cos\sigma \, \, \sech \left(\delta x  \right) + \sin\sigma \, \, \tanh \left( \delta x \right) \right] \, .
\ea
As already pointed out, current-stream interaction driven instabilities may have an important role in the dynamical evolution of several astrophysical plasma structures characterized by a low $\beta$ regime. For instance, the slow wind acceleration region above Sun's helmet streamers is characterized by a typical Alfv\'en speed less than $750~\vunit$~\cite{mann03}, a sound speed of about $100~\vunit$ and a typical  differential velocity of fast and slow streams of about $300-400~\vunit$. For our simulations, we consider the following general settings: the Alfv\'enic Mach is $\approx 0.73$, the sonic Mach number is equal to $3$ and so we have a supersonic and subalfv\'enic current$+$flow system characterized by a $\beta$ of $0.07$. As already pointed out, the angle $\sigma$ defines the initial direction of the AMF relative to the basic flow, having $\sigma = 0$ when the AMF is parallel to the basic velocity field and  $\sigma=\pi/2$ when it is orthogonal to the velocity field. The width of the jet/wake, $a_v$, provides the reference length to set the MHD equations dimensionless, $\tilde{L}$, and hence the dynamic time is $\tilde{t} = a_v / v_a$, while $\delta=a_v / a_b$ measures the ratio between the jet/wake shear width and the current sheet width ($a_b$). We consider the perfect gas equation of state $p_0 = \rho_0\, T_0$ that initially gives 
\ba
\rho_0 & = & \mbox{constant} = 1.0 \\
\label{basic_ffeq}
T_0 & = & c_{0s}^2/\gamma = \mbox{squared sound speed} \, ,
\ea
and we assume a polytropic equation with $\gamma$ equal to 5/3.

The PB basic fields are given by the following relations
\ba
\label{PB_v0y}
v_{0y}(x) & = & \md{M}_a \, \sech^2 \left( x \right) \\
\label{PB_b0y}
B_{0y}(x) & = & \cos\sigma \, \, \tanh \left( \delta x \right) \\
 \label{PB_b0z}
B_{0z}(x) & = & \sin\sigma \, \, \tanh \left( \delta x \right) \, ,
\ea
Pressure equilibrium condition is here assured by a gradient in the temperature profile
\be
\label{basic_ppeq}
T_0 = \frac{1}{2} \, (1 + \beta_{\infty}) - \frac{1}{2} \, \left(B_{0y}^2 + B_{0z}^2 \right)
\ee
where $\beta_{\infty}$ is the plasma beta at the cross-stream boundary for $t = 0$. 

The value of $\delta$ is observed to be a critical parameter for the evolution of the above-defined systems and we assume it equal to $10$. Such quite high value determines a strong gradient in the components of the basic magnetic field and this produces the conditions for an effective instability-triggering mechanism~\citep{dahlburg97,bettarini06}). Besides, from Eqs.~\eqref{FF_v0y}~-~\eqref{FF_b0z} and Eqs.~\eqref{PB_v0y}~-~\eqref{PB_b0z}, we can observe that varying the angle $\sigma$ produces a rotation in the AMF. So, a strong initial magnetic field gradients can affect in a non-trivial way the evolution of the system according to the chosen initial configuration, both in the direction determined by the jet/wake and in the orthogonal one.  

We use two different sets of initial perturbations consisting in a proper two-dimensional space forcing on the cross-stream component of velocity in the jet/wake plane 
\be
\label{perturbation}
v_{1x}(x,y,z) = \varepsilon \, \md{F}(x) \, \sin \left(\frac{2\pi l}{L_y} \, y + \frac{2\pi m}{L_z} \, z + \phi_{r} \right) \, ,
\ee
being $( l , \, m )$ the excited modes in $y$ and $z$ respectively, with $l \in [0, 12]$ and $m \in [-6,6]$, $\phi_{r}$ a random phase, and $\varepsilon$ the initial perturbation amplitude. In~\eqref{perturbation}, we set  
\be
\label{pert_odd}
\md{F}(x) = 2 \, \tanh \left( \delta \, x \right) \, \sech \left( \delta \, x\right)
\ee
or
\be
\label{pert_even}
\md{F}(x) = \sech \left( \delta \, x \right)
\ee
in order to force the system with a function whose parity properties should resemble that of the dominating instability. 

We use a three-dimensional uniform grid with resolution $\left[n_x,n_y\right] \, \otimes \, n_z = (480,120,30)$ with the  following dimensionless sizes 
\ba
\label{box}
L_x & = & \md{D}_x \, [-2\pi , 2\pi] \\
L_y & = & k_{min}^{-1} \, [0 , 2\pi] \\
L_z & = & \md{D}_z  [0 , 2\pi] \, ,
\ea
where $\md{D}_x$ and $\md{D}_z$ are coefficients chosen to properly adapt the size of the grid along the $x$ and $z$ directions, respectively, while $k_{min}$ is the smallest perturbation wave-number along the $y$-direction which set the largest wavelength we can consider. The values of $L_x$ and $N_x$ have been chosen in order to have the cross-stream boundaries located sufficiently far from the current sheet in order to prevent a stabilizing effects from the walls, but at the same time their distance must still allows us to resolve the field gradients within the current sheet with a reasonable number of grid points. A reasonable value of $N_z$ is chosen to follow the dynamics both in the linear and in the nonlinear regime. Higher-resolution test simulations (up to double $N_z$) did not show sensitive differences both in the spectral behavior as well as in the real-space dynamics. For what regards the stream-wise direction, it must be chosen such to allow us to observe in the nonlinear regime both the direct cascade towards higher wave-vectors and a inverse cascade towards large scale by means of a coalescence process.
\bft
\begin{center}
\includegraphics[width=0.45\textwidth]{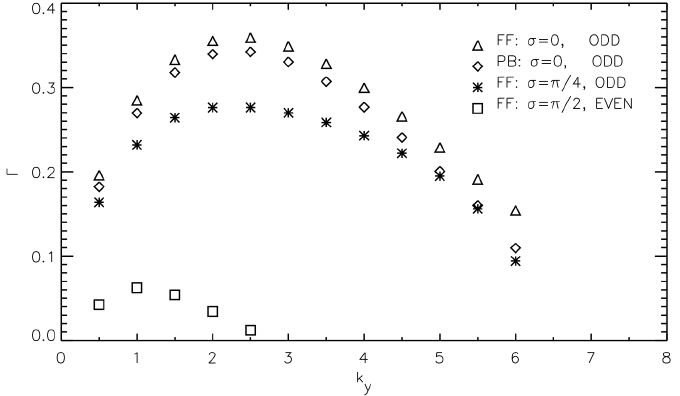}
\caption{Two-dimensional dispersion relations obtained by a linear two-dimensional code for some of the PB and FF cases under investigations. For the FF ``$\pi/2$'' case, the system was perturbed by the even function~\eqref{pert_even} instead of the odd function~\eqref{pert_odd}.}
\label{fig2}
\end{center}
\ef

We verified our choice of the main parameters ($M_a$, $\delta$, $\beta$) by means of a two-dimensional linear code whose details are reported in \citep{landi08}: the maximum linear growth rates correspond to wave-vectors in the range $\left[1, 3\right]$ (see Fig.~\ref{fig2}). We choose hence the following set of parameters for the numerical domain:
\be
\label{boxparam}
\delta = 10 \Rightarrow \left\{
\begin{array}{lll}
\md{D}_x & = 0.75 & \Rightarrow \, L_x = [-3\pi/2 , 3\pi/2] \\ 
k_{min} & = 0.5 & \Rightarrow \, L_y = [0 , 4\pi] \\
\md{D}_z & = 1 & \Rightarrow \, L_z = [0 , 2\pi] \, .
\end{array}
\right .
\ee
So, this stream-wise size implies that the wave-vector corresponding to the above mentioned maximum linear growth rate should corresponds to a wave-number $l$ in a range of $\left[2, 6\right]$. 

The simulations are performed by means of a $7^{th}$ order compact scheme in order to have a low numerical dissipation influencing the expected resistivity-driven linear dynamics, and we consider a Lundquist number equal to $2000$. As in~\citet{landi08}, we verified that the intrinsic numerical dissipation, implicitly introduced by the adopted numerical scheme, is much lower than the explicit diffusivity assumed. In particular, by setting $n_z = 1$ and canceling out the equilibrium magnetic field diffusion, we have verified that the two-dimensional linear evolution in \echo is in very good agreement with the simulations performed by using the linear code.

In the next section we present results of the three-dimensional instability both in the linear (\S~\ref{sec3:1}) and non linear (\S~\ref{sec3:2}) regime. To study the linear regime, as for the two-dimensional simulations, the equilibrium magnetic field diffusion, that affects the growth rates of the linear modes \citep{landi08}, is cancelled out. On the contrary, the study of the non linear regime has been performed by including the effect of the diffusion of the equilibrium magnetic field. In Tab.~\ref{tabgen}, we report the simulations analyzed in the paper.

\section{Results on the $3$D instability} \label{sec3}
\bfh
\begin{center}
\includegraphics[width=0.45\textwidth]{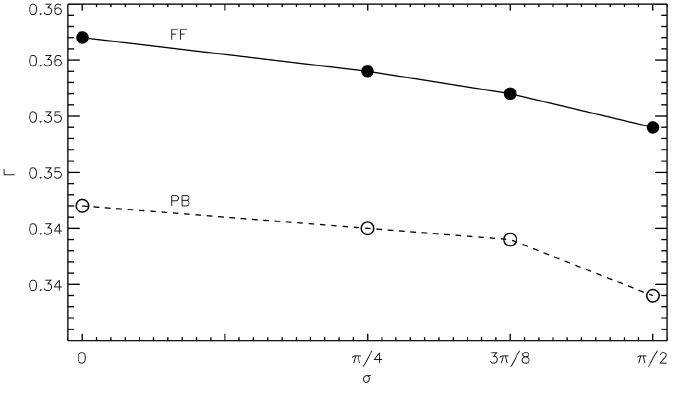}
\caption{Maximum growth rate as function of $\sigma$ for the FF (solid line) and PB (dashed) configurations, during the linear regime of the three-dimensional simulations.}
\label{fig3}
\end{center}
\ef

We consider several values of $\sigma$ for both the FF and PB configuration, also for continuity reasons with the two-dimensional analysis \citep{bettarini06} of these configurations: $0$, $\pi/8$, $\pm \pi/4$, $3\pi/8$, and $\pi/2$. A particular attention must be given to the FF cases with $\sigma = 3\pi/8$ in correspondence of which it is observed a peculiar two-dimensional dynamics characterized by a maximum effect in the asymmetries introduced by magnetic field basic configuration~\citep{bettarini06}. 

\begin{table}
\caption{Summary of the performed three-dimensional simulations. In the first column, we point out the the two possible basic configurations: PB = ``pressure balance'' equilibrium; FF = ``force-free'' equilibrium. In the second column, we report considered values of the angle $\sigma$. For all simulations the following parameters are used: $\delta = 10, \beta \eqsim 0.07, \md{M}_s = 3, \md{M}_a \approx 0.7, \md{S} = 2000$.\label{tabgen}}
\begin{ruledtabular}
\begin{tabular}{lccr}
&basic configuration & $\sigma$ &\\
\hline
&PB, FF & $0$&\\
&PB, FF & $\pi/8$&\\
&PB, FF & $\pm \pi/4$& \\
&PB, FF & $3\pi/8$ &\\
&PB, FF & $\pi/2$ &\\
\end{tabular} 
\end{ruledtabular}
\end{table}

\bts
\begin{center}
\includegraphics[width=0.9\textwidth]{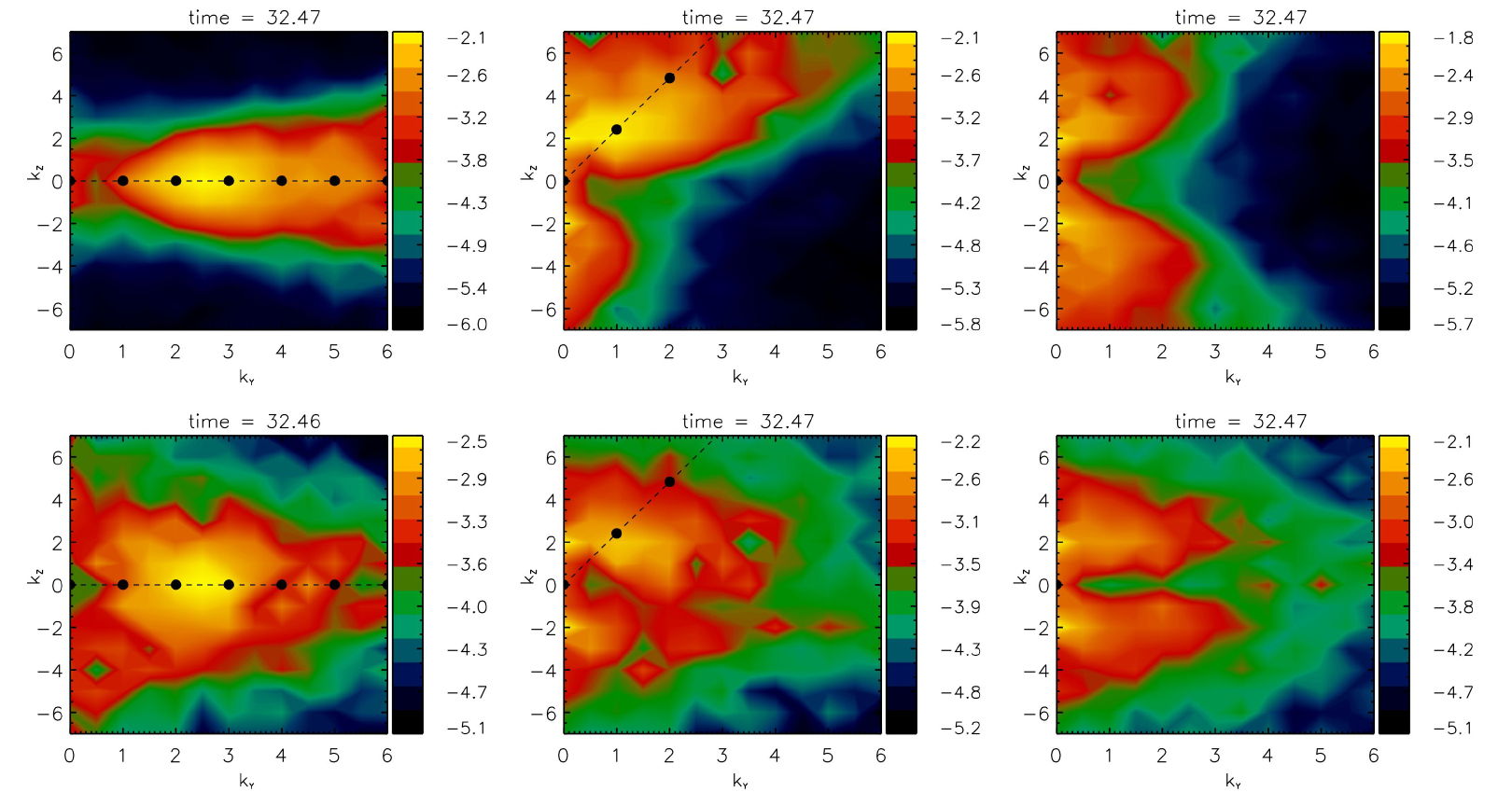}
\caption{(Color online) Magnetic energy (logarithmic scale) in the $k_y- k_z$ Fourier space for three different values of $\sigma$ ($0$ in the left panel, $3\pi/8$ in the mid panel, and $\pi/2$ in the right panel, respectively) for FF (upper row) and  PB (lower row) at the end of the linear regime. The point-dashed line highlights the direction of the AMF.} 
\label{fig4}
\end{center}
\efs

\subsection{Linear regime} \label{sec3:1}

Consider the results from a two-dimensional analysis: For $\sigma < \pi/2$ the varicose-resistive modes  dominate the instability dynamics~\citep{dahlburg97}; a maximum growth rate of about $\Gamma \sim 0.35$ is attained for $\sigma = 0$ decreasing as $\sigma$ increases~\citep{bettarini06}. For $\sigma=\pi/2$, the varicose-resistive modes are no longer unstable: the AMF is now orthogonal to the jet/wake and hence it does not prevent the development of a KH dynamics~\citep{chandrasekhar61,bettarini06}. In Fig.~\ref{fig2}, we show the dispersion relation ($\Gamma$ as function of $k_y$) for a few cases. The $\sigma=\pi/2$ case refers to the sinuous modes given by Eq.~(\ref{pert_even}) and it is evident that the growth rate of the KH instability ($\Gamma\sim 0.07$) is much lower than those for the varicose-resistive modes with $\sigma<\pi/2$.

In the three-dimensional simulations, the situation changes significantly. In fact, as shown in Fig.~\ref{fig3}, we obtain the same growth rates of about $\Gamma \approx 0.35$, regardless of the value of $\sigma$. Moreover, all the cases are characterized by the development of varicose-resistive modes, as we can infer from the fact that the growth rates match the values observed in the two-dimensional linear simulations with odd perturbations. In general, the duration of the linear phase is slightly longer in the PB cases than in FF's ones since the growth rates of the fastest modes are a bit smaller. This is consistent with the 2D linear simulations. This different behavior is clearly related by the fact that in three dimensions all instability modes are allowed to developed. In Fig.~\ref{fig4}, we report the magnetic energy spectrum in the Fourier space $k_z-k_y$ at a given time during the linear regime for the FF (upper row) and the PB (lower row) configurations: the cases with $\sigma = 0, 3\pi/8$ and $\pi/2$ are shown in the left, middle, and right column, respectively. As the instability dynamics sets in, both the FF and the PB configurations where the AMF is aligned with the basic fluid jet ($\sigma = 0$) are driven mainly by  the modes characterized by $k_z = 0$, although a small amount of energy is present in modes with  $k_z = \pm 1, 2$, at least for the PB configuration. Consistently with the linear code results, The observed fastest-growing linear mode lies between $k_y = 2, 2.5$.
\bts
\begin{center}
\includegraphics[width=0.9\textwidth]{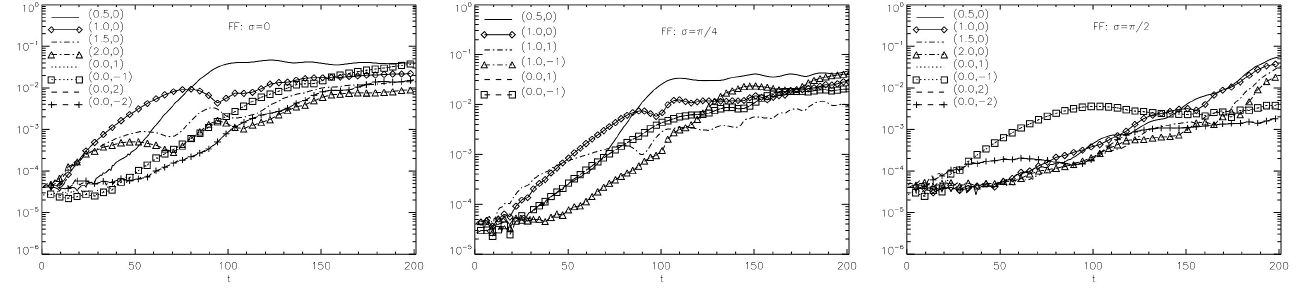}
\caption{From left to right, the most relevant Fourier modes of the magnetic energy in the nonlinear phase as a function of time for the FF configurations with $\sigma = 0$ (left), $\sigma=\pi/4$ (middle), and $\sigma=\pi/2$ (right).}
\label{fig5}
\end{center}
\efs
If now we consider a value of $\sigma$ between $0$ and $\pi/2$, such as $\sigma = 3\pi/8$ shown in the middle panel of Fig.~\ref{fig4}, we observe that the most unstable modes are no more aligned with the jet/wake flow. The instability grows along a preferential direction that is selected by the direction of the AMF, underlined in the figure by the dash-dotted line. The most unstable modes for this case are those characterized by $k_y = 1$ and $k_z = 2, 3$, corresponding to $|k|$ in the range $[\sqrt{5},\sqrt{10}]$ which is consistent with the values predicted by the linear simulations for $\sigma = 0$. Same results, not shown here, are obtained for $\sigma = \pi/8$, and $\sigma=\pi/4$. Moreover, the same results is obtained in the case of $\sigma = -\pi/4$ (FF) configurations: this implies that the way the cross component of the magnetic field rotates through
the current sheet is not relevant for the dynamics of the system. 

As shown previously, in the two-dimensional case, for $\sigma=\pi/2$, the system should be unstable under KH like instability. However, if we look at the right panel of Fig.~\ref{fig4}, we note that both FF and PE configurations are dominated by modes aligned with the $z$-direction, the direction of the AMF field, although a significant dose of energy is still present in modes with $k_z=0$ for the PB case. The most unstable modes lie in the wave-vector range $k_z \approx \pm 2-3$: the system is dominated by resistive-varicose instabilities. The presence of excited modes along the $y$-direction is due to the presence of the KH instability, stronger for the PB case because of the absence of the stabilizing effect of the perpendicular magnetic field inside the current-sheet.

In conclusion, regardless of the $\sigma$ values, the most unstable modes driving the system throughout the linear regime are resistive-varicose and they are selected according to the AMF direction regardless of the detailed structure of the magnetic field within the current-sheet, which is the type of equilibrium under consideration. In particular, these modes are characterized by the condition $k \times \vect{B}(|\delta x| \gg 1) = 0$. Such preferential direction determined by the asymptotic magnetic field in the linear evolution of the instability was observed also in the case of the incompressible and compressible three-dimensional tearing mode evolution by~\citet{onofri04} and~\citet{landi08} respectively.
\bts
\begin{center}
\includegraphics[width=0.9\textwidth]{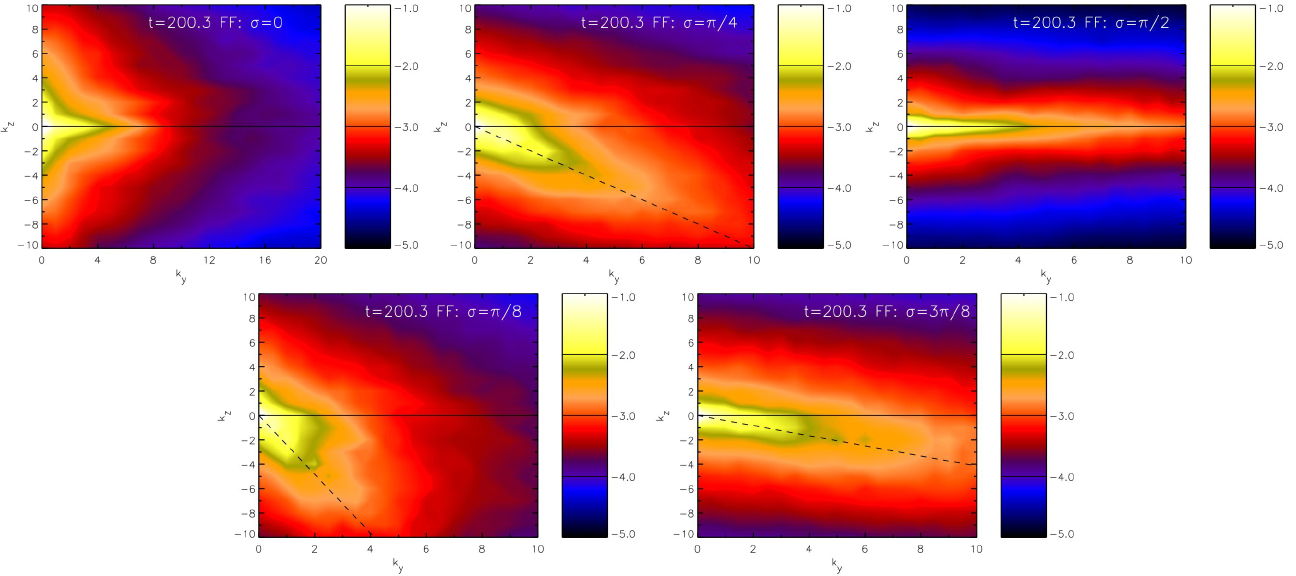}
\caption{(Color online) Magnetic energy in the $(k_y-k_z)$ Fourier space for the FF configurations corresponding to different values of $\sigma$ in the well-developed nonlinear regime. The dashed line highlights the $\sigma-\pi/2$ direction.}
\label{fig6}
\end{center}
\efs
\bft
\includegraphics[width=0.45\textwidth]{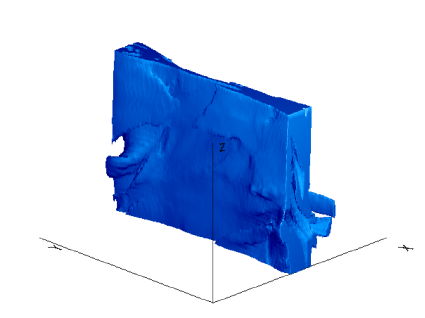}
\caption{(Color online) Three-dimensional plasma pressure iso-contour at $t = 200$ for the FF configuration with $\sigma = 0$. The shaded surface encodes the region of space where the pressure is more than $65$\% of its maximum.}
\label{fig7}
\ef

\subsection{Nonlinear regime} \label{sec3:2}

Although the nonlinear regime of the current (neutral) double vortex sheet is characterized by a complex behavior which depends on the initial evolution of the most unstable modes, it is possible to identify some key-features. 

\subsubsection{FF cases} \label{sec3:2:1}
\bts
\begin{center}
\includegraphics[width=0.9\textwidth]{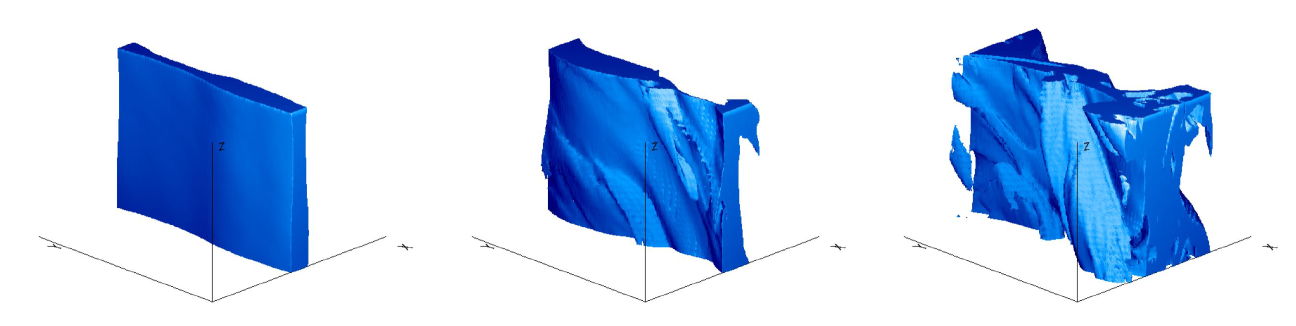}
\caption{(Color online) Three-dimensional plasma pressure iso-contour at three different instants during the nonlinear evolution of the FF configuration with $\sigma=\pi/4$. These images refer to $t = 100$ (left panel), $t = 150$ (middle panel) and $t = 200$ (right panel). The shaded surface encodes the region of space where the pressure is more than $50$\% of its maximum.}
\label{fig8}
\end{center}
\efs

We start from Fig.~\ref{fig5} to consider all the cases determined by a different value of $\sigma$. Here, we show the temporal evolution of some magnetic energy modes for three different FF configurations: $\sigma=0$ (left panel), $\sigma = \pi/4$ (middle panel) and $\sigma=\pi/2$ (right panel).

\begin{description}
\item[$\sigma = 0$:] From about $t \sim 50$, it is clearly observed a coalescence process driving the system to the maximum length scale allowed by our numerical domain. In fact, the inverse cascade ends as soon as the largest wave-number mode dominates at about $t \sim 80$. In the meanwhile, the orthogonal modes (those with $k_y = 0$ and $k_z \neq 0$) keeps on growing and, from $t \sim 150$ on, they start to determine the structure of the system. At about  $t\sim 200$, most of the energy is symmetrically distributed in the ``secondary'' (orthogonal) modes as well as in the first primary modes ($k_y \neq 0$ and $k_z = 0$), as shown in the top-left panel of Fig.~\ref{fig6} where the magnetic energy spectrum in the plane $(k_y, k_z)$ in the well-developed nonlinear regime is shown for different $\sigma$-valued FF initial configurations. This dynamics is analogous to those shown in~\citet{landi08} analyzing the three-dimensional compressive tearing evolution. Also in that study, the nonlinear magnetic energy spectrum is characterized by two preferred directions resulting in a system modulated in two different ways in the physical space. In Fig.~\ref{fig7} it   is shown the three-dimensional plasma pressure iso-contour for the "0" case {\bf at $t=200$}  and it is possible to observe a coalesced plasma structure along the stream-wise direction modulated along the span-wise axis.
\bts
\begin{center}
\includegraphics[width=0.65\textwidth]{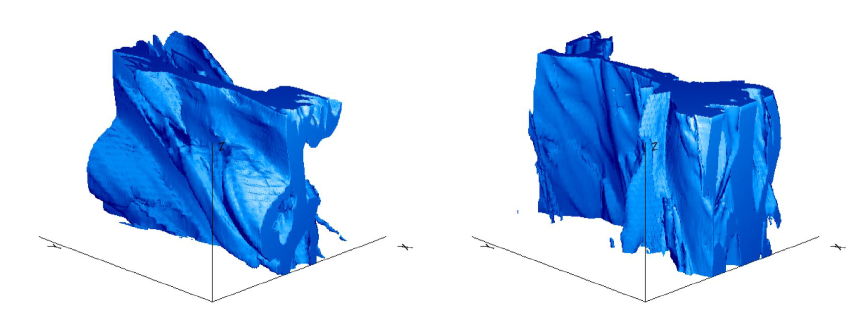}
\caption{(Color online) Three-dimensional plasma pressure iso-contour at the end of the simulation ($t = 200$) for two different FF configurations: $\sigma=\pi/8$ (left panel) and $\sigma = 3\pi/8$ (right panel). The shaded surface encodes the region of space where the pressure is more than $55$\% of its maximum.}
\label{fig9}
\end{center}
\efs
\begin{figure*}
\includegraphics[width=0.65\textwidth]{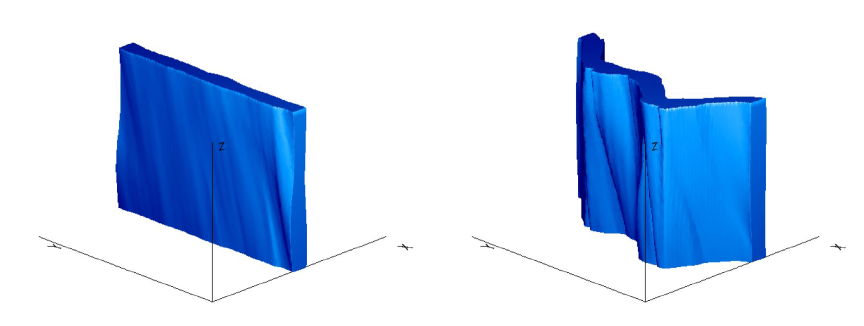}
\caption{(Color online) Three-dimensional plasma pressure iso-contours of the FF configuration with $\sigma=\pi/2$. Image refers to $t=125$ (left panel) and $t = 200$ (right panel). The shaded surface encodes the region of  space where the pressure is more than $50$\% of its maximum.}
\label{fig10}
\end{figure*}
\bts
\begin{center}
\includegraphics[width=0.9\textwidth]{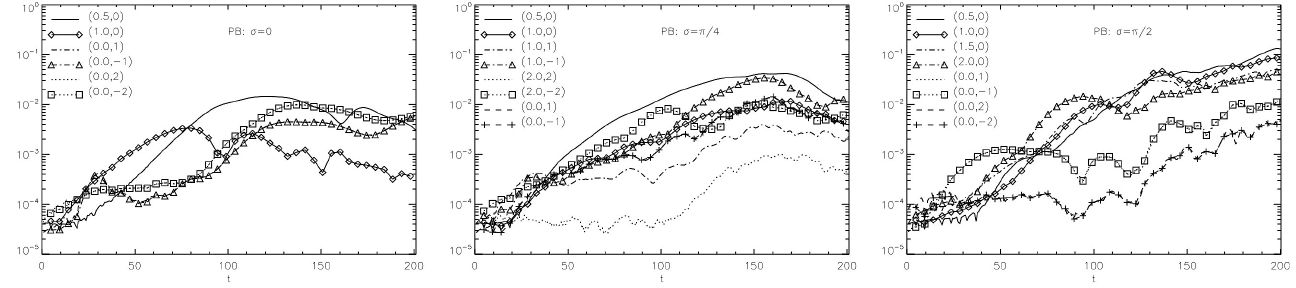}
\caption{From left to right, the most relevant Fourier modes of the magnetic energy in the nonlinear phase as a function of time for the PB configurations with $\sigma = 0$ (left), $\sigma=\pi/4$ (middle), and $\sigma=\pi/2$ (right).}
\label{fig11}
\end{center}
\efs

\item[$0 < \sigma < \pi/2$:] The overall dynamics is not very different from the previous case, provided that the nonlinear driving modes are properly considered. For instance, in the ``$\pi/4$'' case shown in Fig.~\ref{fig5} (middle panel), we observe a coalescence process with about the same time-scale of the ``0'' case. Afterwards, the modes with a given direction to the AMF starts to increase, in particular the mode $(1, -1)$, and in the far nonlinear regime they own most of the system energy as shown in the top-middle panel of Fig.~\ref{fig6}. In the physical space, the rise up of the secondary instability determines that the globally-coalesced structure observed in the left panel of Fig.~\ref{fig8} at about $t \sim 100$, at later times is modulated along the direction located at $\pi/4$ with respect the $x-y$ plane as shown in the middle and right panels of Fig.~\ref{fig8}. In Fig.~\ref{fig6} we show also the magnetic energy spectra at the end of the nonlinear regime also for the cases $\sigma = \pi/8$ (bottom-left panel) and $3\pi/8$ (bottom-right panel). The dashed lines represent the $\sigma-\pi/2$ direction and they well fit the observed spectrum anisotropy direction. In all cases, It is worth pointing out that this privileged direction at about $t \sim 200$ is orthogonal to the direction of the most energetic modes in the linear regime (see Fig.~\ref{fig2} as reference for the $3\pi/8$ case). The three-dimensional plasma pressure iso-contours at the end of the simulations for the $\sigma=\pi/8$ (left panel) and for $\sigma=3\pi/8$ case (right panel) shown in Fig.~\ref{fig9} reveal the effects in the physical space of the behavior observed in the magnetic energy spectra. 

\item[$\sigma = \pi/2$:] As for the previous cases, the first part of the non linear evolution is characterized by the growth of the primary instability along the AMF, that is now the span-wise direction $z$. At about $t=125$, the largest coalesced structure with respect to our numerical box in the $z$ direction is developed. As shown in the left panel of Fig.~\ref{fig10}, in the physical space enhancements of the plasma pressure underlines the presence of magnetic islands, as it is typical of resistive driven instabilities. Yet, the modes orthogonal to the primary ones are still growing and they reach the same energy level at about $t = 150$. So far, system's behavior is mostly analogous to the previously described cases, characterized all by an energy density almost equally distributed both in primary and secondary modes. Yet, the final part of the nonlinear evolution presents a completely different dynamics: the modes along the stream-wise direction are dominating the others of about one order of magnitude in the energy spectrum (see Fig.~\ref{fig5}~and~\ref{fig6}). An analysis in the physical space reveals that the prevailing modes have a different nature with respect the previous cases. In fact, in the right panel of Fig.~\ref{fig10}, we observe that at $t = 200$ the plasma pressure shows the typical sinuous profile of KH instabilities. So, although it is completely overwhelmed by the resistive-varicose mode in the linear stage, nevertheless the KH dynamics appears to dominate in the well-developed nonlinear regime and as it seems to be mainly driven by the presence of the velocity shear, we expect it to influence strongly the wake (jet) acceleration (deceleration) (see in section~\ref{sec3:3}).
\end{description}

So, as already mentioned, we observe a qualitatively similar behavior for different values of $\sigma < \pi/2$. A resistive linear regime is driven by the modes selected according to the direction of the equilibrium asymptotic magnetic field initially defined in the $(y, z)$ plane. The nonlinear regime is initially characterized by the formation of magnetic islands then undergoing to a coalescence process. Afterwards, the onset of secondary instabilities kink the coalesced structures: in the later stages of the simulations, we observe plasma pressure enhancements strongly modulated by modes orthogonal to the primary modes, as similarly observed in three-dimensional tearing simulations~\citep{landi08}. In the Fourier space, the presence of the secondary modes is evident in the magnetic density energy contained in modes orthogonal to the primary ones. For $\sigma = \pi/2$, the system is also unstable to KH instabilities and this changes significantly its nonlinear dynamical path. In this case, the later stage of system's evolution is dominated by sinuous modes.

\subsubsection{PB cases} \label{sec3:2:2}

Now, we consider the nonlinear evolution of the PB equilibrium configurations for different values of $\sigma$. In Fig.~\ref{fig11}, it is shown the time evolution of the most important magnetic energy modes for the cases $\sigma=0$, $\sigma=\pi/4$, and $\sigma=\pi/2$.

\begin{description}
\item[$\sigma = 0$:]  We observe that at the beginning of the nonlinear regime the system is still driven by one of the linearly-dominating modes, $(k_y,k_z)=(1,0)$ leading it to be structured almost on the largest length-scale. In fact, after a few dozens of time steps we observe a inverse cascade towards the maximum length scale allowed by our numerical box, $(0.5,0)$, followed by a saturation plateau of the nonlinear regime lasting up to about $t \sim 130$. In the meantime, the orthogonal modes $(0,\pm 1)$ and $(0,\pm 2)$ grow as well. So, the two-dimensional structured system starts to develop in the $z$-direction according to these modes which, together with their harmonics, drive the system till the end of the simulation. Primary modes directed along the $y$-direction, except for the fundamental one $(0.5,0)$, differently from the FF configuration, are strongly damped: This is evident, for example,  by comparing the mode $(1.0,0)$ in Fig.~\ref{fig11}, left panel, with  the analogous one for the FF configuration in Fig.~\ref{fig5}. So, from an early coalescing structure, the system presents at the end of the simulation an essentially two-dimensional configuration in the $(x,z)$, as observed in the plasma pressure structure shown in Fig.~\ref{fig12}. This behavior is confirmed also by the magnetic energy spectrum at the end of the simulation, shown in the left panel of Fig.~\ref{fig13}: most of the magnetic energy is now contained in modes directed along the $z$-direction and we do not observe the energy redistribution between primary and secondary modes as in the analogous FF configuration. This result is consistent with those obtained in the pure three-dimensional tearing instability simulations for PB configuration~\citep{landi08}.
\bts
\begin{center}
\includegraphics[width=0.9\textwidth]{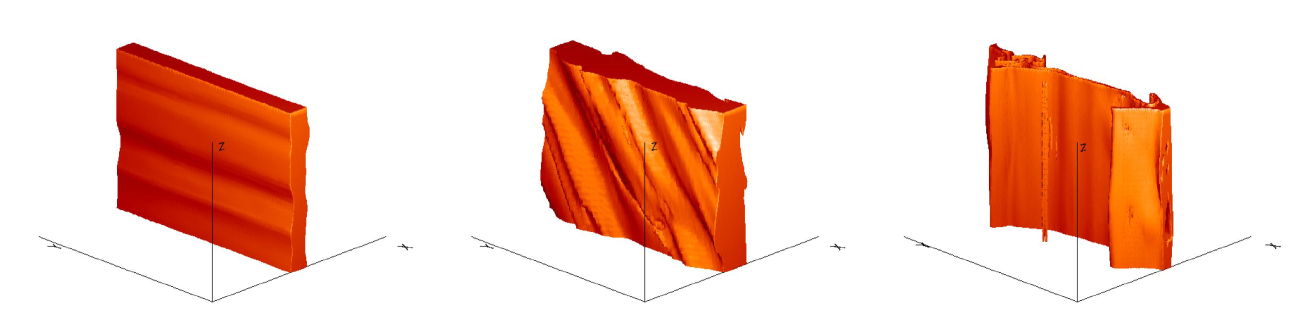}
\caption{(Color online) Three-dimensional plasma pressure iso-contour during the nonlinear evolution of the PB configuration with $\sigma=0$ (left), $\sigma=\pi/4$ (middle) and $\sigma=\pi/2$ (right). The shaded surface encodes the region of space where the pressure is more than $60$\% of its maximum. For $\sigma=0$ and $\sigma=\pi/4$ the pressure is evaluated at $t=200$, while for the $\sigma=\pi/2$ the reference time is $t=150$.}
\label{fig12}
\end{center}
\efs
\bts
\begin{center}
\includegraphics[width=0.9\textwidth]{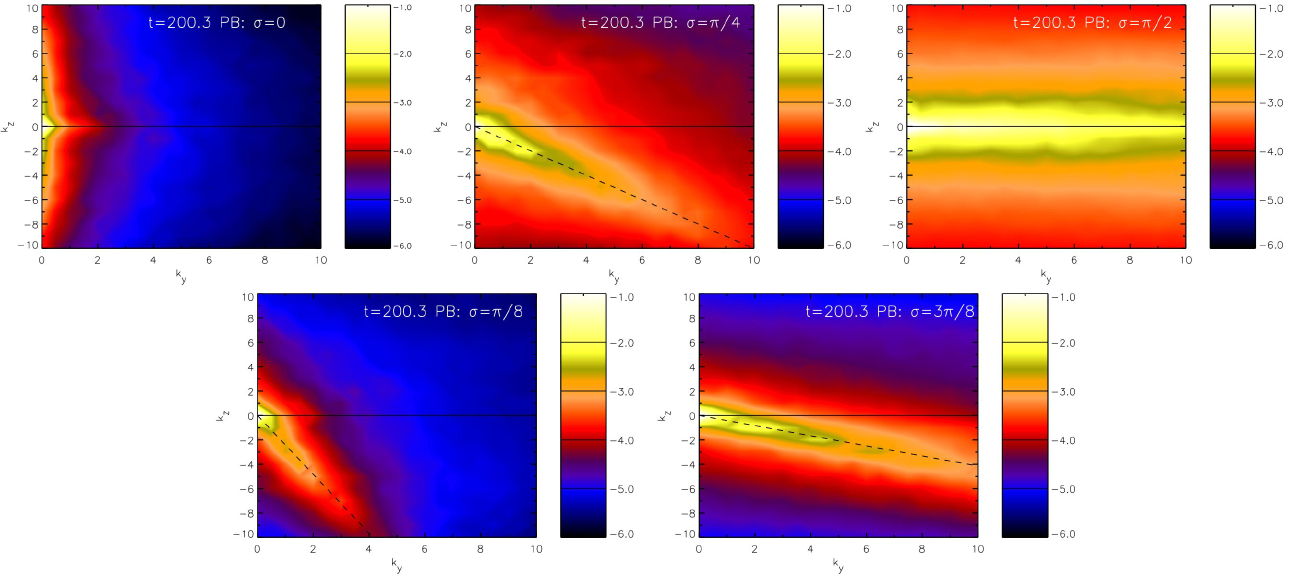}
\end{center}
\caption{(Color online) Magnetic energy in the $(k_z-k_y)$ Fourier space for the PB configurations with different values of $\sigma$ at $t\sim 200$. As for the FF case (Fig.~\ref{fig6}), the dashed line highlights the angle $\sigma-\pi/2$.}
\label{fig13}
\efs

\item[$0 < \sigma < \pi/2$:] Let us focus on one emblematic case, such as for example $\sigma=\pi/4$. After the first phase, where the system evolution is determined by the primary three-dimensional modes characterizing the linear regime, as shown in the middle panel of Fig.~\ref{fig11}, we have an increase of the contribution of the mode $(k_y,k_z) = (0.5,0)$ which drives the dynamics till the end of the simulation. At the same time, modes orthogonal to the primary ones increase in time such that, at the end of the simulation, they contain a large portion of the magnetic fluctuating energy, as it is shown in the middle panel of Fig.~\ref{fig13}. Again, the dashed line locates the $\sigma-\pi/2$ direction and it well fits the observed spectrum anisotropy direction pointing out that the privileged direction at about $t \sim 200$ is orthogonal to the direction of the most energetic modes in the linear regime. A similar behavior is observed in the magnetic energy spectrum for the "$\pi/8$" and "$3\pi/8$" cases (bottom-left and bottom-right panels of Fig.~\ref{fig13}, respectively). The physical behavior of the system is shown in Fig.~\ref{fig12} (middle panel), where it is evident the instability modulation of the whole structure along the preferred direction. 

\item[$\sigma = \pi/2$:] The overall behavior is similar to the corresponding FF configuration, although it appears faster in its evolution. At $t = 50$, it is attained the saturation in the magnetic energy power in the smallest span-wise ($z$) wave-vector and it is observed a rapid growth of the energy of stream-wise ($y$) modes. At $t=100$, the system is strongly structured in the $(x,y)$ plane with a typical varicose-mode configuration. However, the presence of KH instability modes leads the system to be structured in a typical sinuous configuration already at $t = 150$ as it is possible to observe in the right panel of Fig.~\ref{fig12}. At the end of the simulation, most of the magnetic energy is contained in stream-wise ($y$) wave-vectors (Fig.~\ref{fig13}, top right panel).
\end{description}

\subsection{Jet/Wake evolution} \label{sec3:3}
In general, a net acceleration effect along the stream-wise direction of the jet/wake embedding the current sheet is observed, while  the mean (positive and negative) velocity contributions along the other directions are negligible for almost all the nonlinear regime. We can measure the net acceleration effect by averaging out the span-wise and stream-wise dependence of the stream-wise component of the velocity field, i.~e. by looking at the function 
\be
\label{eq_vy_yz}
\langle v_y \rangle_{yz}=\frac{1}{L_y L_z} \int_0^{L_y}\int_{0}^{L_z} v_y(x,y,z) \, dy dz \, .
\end{equation}
In Fig.~\ref{fig14}, we plot this quantity for different values of the angle $\sigma$ and for both FF and PB equilibria (first and second panel). For comparison, we report the same profiles for two-dimensional simulations with the same initial condition in the well-developed nonlinear regime (third and fourth panel). It is manifest the strong differences from the two- and the three-dimensional case. Even though we are exploring a slightly different parameter space from~\citet{bettarini06}, the two-dimensional simulations we performed confirms the overall dynamics there discussed.  In three-dimensions, we observe a weaker acceleration effect and a less pronounced enlargement of the wake, both in the FF and PB cases. Because of the insurgence of the secondary instability, dips in the velocity profiles observed in two-dimensional simulations, caused by the presence of coherent vortices structures inside the current-sheet region, are no longer present. 
\bts
\begin{center}
\includegraphics[width=0.9\textwidth]{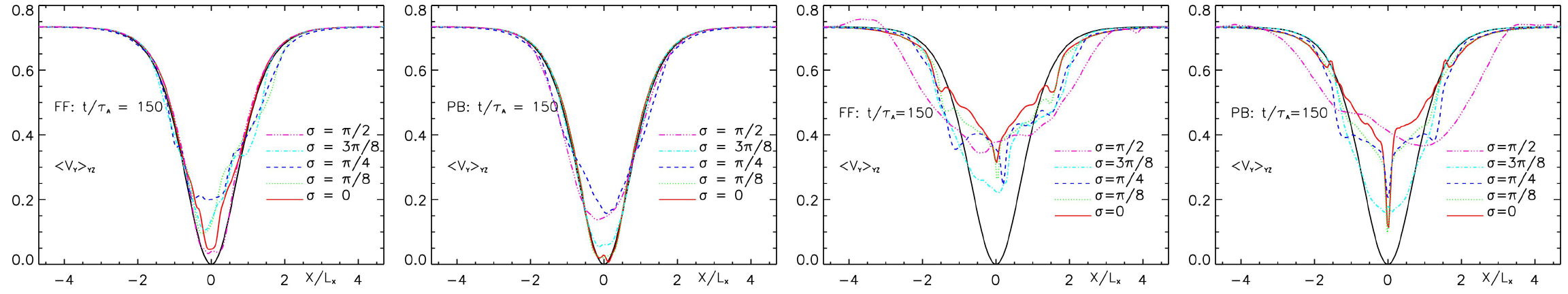}
\caption{(Color online) Wake profile of the stream-wise component of the velocity, as computed using Eq.~\ref{eq_vy_yz}, for all the values of $\sigma$. The black solid line in all plots is the wake profile a $t = 0$ shown as reference. The first and second panels refer respectively to the FF and PB three-dimensional cases (see section~\ref{sec3:3}), whereas, for comparison, in the third and fourth panels it is reported the FF and PB two-dimensional cases respectively.} 
\label{fig14}
\end{center}
\efs

\section{Discussions and conclusions} \label{sec4}

We consider the fully three-dimensional linear and nonlinear evolution of a plane neutral/current double vortex sheet consisting in a neutral/current-sheet embedded in a sheared supersonic flow. Several parameters can influence the evolution of the initially perturbed system: The initial equilibrium configuration, the magnetic and velocity fields relative geometry, the plasma $\beta$, the sonic and Alfv\`enic Mach numbers. Here we have considered the typical values describing jet/wake flows in the solar active regions at a few solar radii: in particular, the configuration can be applied to the wake model of the slow component of the solar wind~\citep{einaudi01,rappazzo05,bettarini06}.

Two different equilibrium configurations have been compared: a force-free (FF) magnetic field whose polarity reversal determines the formation of a current sheet inside the plasma flow and a pressure balance (PB) magnetic field where the polarity reversal is obtained by means of a neutral plane inside the flow. Differently from the 2D simulations~\citep{bettarini06}, for all orientations of the asymptotic magnetic field, both the FF and PB equilibrium configurations are dominated by varicose-resistive modes. The most unstable ones are characterized by wave-vectors whose direction is parallel to the asymptotic magnetic field. The presence of a rotational component inside the current sheet, as for the FF case, does not affect significantly the linear behavior of the system. Such selection rules has been already observed in previous analysis of three-dimensional tearing mode evolution~\citep{landi08}.

The nonlinear regime is always characterized by a two-stage evolution: for $\sigma<\pi/2$, the varicose-resistive modes (primary modes) start to coalesce and then a secondary instability develops. In the physical space, it results in the formation of magnetic islands and pressure enhancement regions which afterwards are deformed by kink-like instabilities. The appearance of the secondary instability is accompanied by the growth of fluctuations whose wave-vector are orotogonal to the primary ones. The resulting magnetic energy spectrum, as for the simple current-sheet instability~\citep{onofri04,landi08}, results thus strongly anisotropic. As pointed out by \citet{landi08}, the presence of the magnetic field inside the current sheet in the FF configuration can reduce the effects of the secondary instability: as a consequence, in the fully developed non linear regime a large dose of magnetic energy is in both the primary and secondary (orthogonal) modes. Because of the absence of a guide field, which has a stabilizing effect on the current-sheet evolution~\citep{onofri04,dahlburg05,landi08}, for the PB configuration, the transition to the secondary instability appears earlier and the magnetic energy spectrum is dominated by these modes. As in \citet{bettarini06} the production of magnetic islands is accompanied by vortices formation which are considered the responsible of the wake (jet) acceleration (deceleration) \citep{einaudi99,einaudi01}. Similarly to what observed by \citet{einaudi99}, the onset of the secondary instability determines the destabilization of the islands, and hence it reduces the acceleration/deceleration effect. 

For $\sigma=\pi/2$, for sonic and Alfv\'enic Mach numbers typical of the heliospheric environment, the system results to be unstable both by tearing and Kelvin-Helmholtz instabilities. With the relatively low Lundquist number here adopted, it results that, differently from the two-dimensional case~\citep{bettarini06}, in three dimensions, the linear regime is dominated by resistive instabilities. However, during the non linear regime, because of the dissipation of the underlying equilibrium magnetic field, the width of the current-sheet increases and, as consequence, there is a decrease in time of the tearing mode growth rate as it goes as $\delta^{3/2}$~\citep{priest00}. As a consequence,  if the diffusion acts for a sufficiently long time, the flow driven instability is able to dominate the system during the non linear regime. Moreover, since the growth rate of the resistive mode scales as $\md{S}^{-1/2}$, it is expected that, for a magnetic configuration where the asymptotic magnetic field is orthogonal to the jet/wake, Kelvin-Helmholtz instabilities will be the prominent dynamics driving the overall system evolution. 
\bts
\begin{center}
\includegraphics[width=0.45\textwidth]{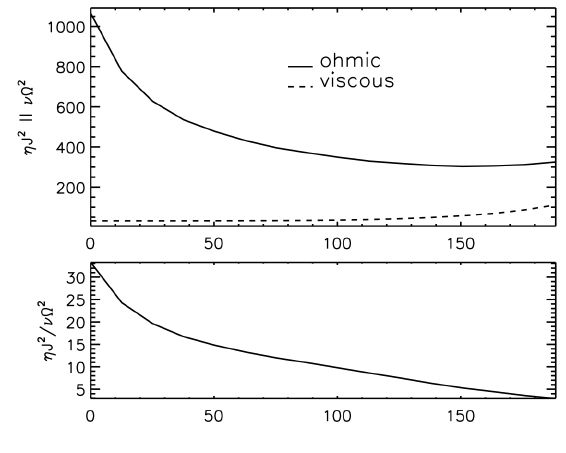}
\caption{(Color online) Top panel: Volume average of the ohmic and viscous dissipation (solid and dashed line, respectively) as a function of the simulation time for the FF simulation with $\sigma = \pi/4$. Bottom panel: Ratio of these quantities as a function of the simulation time.} 
\label{fig15}
\end{center}
\efs
In the present work the current (neutral) double vortex sheet evolution is investigated in the limit of zero kinematic viscosity, in analogy with~\citep{landi08}. In fact, although the influence of viscous effects can be retained with respect to the resistivity-driven dynamics, however it has been suggested that the overall evolution depends essentially on the Hartmann number which gives a measure of the relative importance of drag forces resulting from magnetic diffusivity and kinematic viscous forces~\citep{dahlburg83}. Fig.~\ref{fig15} shows an \emph{a posteriori} computation of the volumetric average of the ohmic and the viscous dissipation (solid and dashed line respectively, in the top panel) and their ratio as a function of the simulation time. Here, a FF simulation with $\sigma = \pi/4$ is considered, but this case is a paradigma for all simulations. The above mentioned quantities are defined respectively as
\ba
\label{ohmic}
\md{W}_{ohm} (t) & = & \int_V dV \, \eta \, |\vect{J}|^2 \\
\label{viscous}
\md{W}_{vis} (t) & = & \int_V dV \, \nu \, |\Omega|^2 \, ,
\ea
where we assume $\nu = \eta$, $\Omega$ is the vorticity vector, $\bnabla \times \vect{v}$, and $V$ is the volume of our three-dimensional numerical box. In general, the overall ohmic dissipation is far bigger than the viscous one: as the motion gets turbulent and vortices form and grow, the contribution of the viscosity becomes more and more important up to be one fifth of the other. Yet, the overall dynamics throughout the linear and nonlinear regime is not affected by that, in particular with regards to the onset of the coalescence process or the triggering of ideal instabilities as the secondary ones~\citep{dahlburg01}. Nevertheless, a further investigation on viscosity effects in longer simulations is mandatory. For several heliospheric and astrophysical environment applications it will be suitable to consider the effect of the radial expansion and/or the presence of an underlying stratified medium:  they can influence not only the onset of the primary instability, but also the length- and time-scales associated with the insurgence of the secondary instability. 

\begin{acknowledgments}
The authors are grateful to Luca Del Zanna and Andrea Verdini for the useful comments and suggestions. Numerical computations  has been performed using the $512$ processor IBM SP5  of the CINECA consortium available through the INAF-CINECA agreement $2006-2007$: High Performance Computing Resources for Astronomy and Astrophysics. M. Velli was supported in part by the NASA LWS TR\&T. This research was supported in part by ASI contract n. I/015/07/0 ``Solar System Exploration''. We are also grateful to the anonymous referee for comments and suggestions.
\end{acknowledgments}



\end{document}